\documentclass{elsart}

\usepackage{graphicx,amssymb,lineno}

\begin{document}

\begin{frontmatter}

\title{The production of Tsallis entropy in the limit of weak chaos 
and a new indicator of chaoticity}

\author[uoa,keaem]{G. Lukes-Gerakopoulos\corauthref{cor}\thanksref{scol}},
\corauth[cor]{Corresponding author.}
\thanks[scol]{Supported by the Greek Foundation of State Scholarships (IKY)
and by the Research Committee of the Academy of Athens.}
\ead{gglukes@phys.uoa.gr}
\author[keaem]{N. Voglis$^\dagger$},
\author[keaem]{C. Efthymiopoulos}
\ead{cefthim@academyofathens.gr}

\address[uoa]{University of Athens,Department of Physics, Section of Astrophysics,
 Astronomy and Mechanics}
\address[keaem]{Academy of Athens, Research Center for Astronomy and Applied Mathematics,
Soranou Efesiou 4, GR-11527, Athens, GREECE}

\begin{abstract}{\small
We study the connection between the appearance of a `metastable' behavior of 
weakly chaotic orbits, characterized by a constant rate of increase of the 
Tsallis q-entropy [J. of Stat. Phys. Vol. 52  (1988)], and the solutions of the variational equations 
of motion for the same orbits. We demonstrate that the variational equations 
yield transient solutions, lasting for long time intervals, during which the 
length of deviation vectors of nearby orbits grows in time almost as a power-law. 
The associated power exponent can be simply related to the entropic exponent for 
which the q-entropy exhibits a constant rate of increase. This analysis 
leads to the definition of a new sensitive indicator distinguishing regular 
from weakly chaotic orbits, that we call `Average Power Law Exponent' (APLE). 
We compare the APLE with other established indicators of the literature. 
In particular, we give examples of application of the APLE in a) a thin 
separatrix layer of the standard map, b) the stickiness region around an 
island of stability in the same map, and c) the web of resonances of a 
4D symplectic map. In all these cases we identify weakly chaotic 
orbits exhibiting the `metastable' behavior associated with the Tsallis 
q-entropy. } 

\end{abstract}


\end{frontmatter}

\section{Introduction}

The usefulness of the so-called `non-extensive q-entropy' \cite{Tsallis01} in 
characterizing the statistical mechanical properties of nonlinear dynamical 
systems has so far been demonstrated in a number of instructive examples 
in the literature (see \cite{Tsallis03} for a comprehensive review). 
In the present paper we focus on one particular property of the Tsallis 
q-entropy, first reported in \cite{Tsallis02,Costa01},
and further explored in \cite{Lyra01,Baranger01,Latora02}.
These authors demonstrated that when a nonlinear dynamical system is in 
the regime of the so-called `edge of chaos' the rate of increase of the q-entropy 
remains constant for a quite long time interval. Tsallis and the coauthors 
\cite{Tsallis02} argued that this behavior of the q-entropy can be 
connected to the phenomenon of a power-law rather than exponential sensitivity 
of the orbits on the initial conditions. In the case of the Feigenbaum attractor 
such a power law was observed by Grassberger and Scheunert \cite{Gras01}.
Grassberger recently \cite{Gras02} questioned the meaning of the q-entropy in
that particular case, but his arguments were convincingly rebutted in \cite{Robledo01}.
Furthermore, in the case of conservative systems, Baranger \cite{Baranger01} found
a constant rate of increase of the usual Boltzmann - Gibbs (BG) entropy in strongly 
chaotic systems such as a generalized cat map or the Chirikov standard map for high 
values of the non-linearity parameter $K$. However, when $K$ is small, there is again a 
transient interval of time in which the q-entropy,  rather than the 
Boltzmann - Gibbs entropy, exhibits a constant rate of increase. 
This phenomenon was found numerically in low-dimensional mappings when $q$ takes
a particular value (in the standard map $q\simeq 0.3$ for $K$ close to $K_c=0.97...$
\cite{Baldovin01,Baldovin02}, while $q\simeq 0.1$ when $K\sim 10^{-1}$ \cite{Ananos01},
and it was called a `metastable state' \cite{Baldovin01}.

The above calculations were based essentially on a `box counting' method. 
Namely, the phase space is divided into a number of cells, and the average 
covering of these cells is found over `many histories' \cite{Baldovin01}, i.e., over 
large ensembles of orbits, when the initial conditions are taken 
inside a very small domain (e.g. a box of size $10^{-2})$. The sensitivity to 
the initial conditions was checked by following nearby orbits with a small initial separation,
(e.g. $10^{-9}$ or $10^{-12}$). Finally, the value of $q$ 
for which the q-entropy grows linearly in time was found by inspection, i.e., 
by trying many different values of $q$. 

On the other hand, the same type of `metastable' behavior 
(as implied by Tsallis in \cite{Tsallis02})
should be recovered more simply if, instead of taking averages over 
many orbits, one calculates the time evolution of the {\it deviation vectors} 
$\xi(t)$ given by solving the variational equations, together with the 
equations of motion, for the orbits inside a weakly chaotic domain. In the 
present paper we consider, precisely, the question of how can the `metastable 
behavior'associated with a constant rate of production of the Tsallis q-entropy 
be justified theoretically by an analysis of the behavior of the variational 
equations in the limit of weak chaos. This analysis is further substantiated by 
numerical results. In fact, from the numerical point of view the method of 
variational equations is advantageous over the method of `many histories' in 
that, when the phase space is compact, the variational equations yield the local 
rate of growth of deviations along one orbit for an arbitrarily long integration 
time, while a calculation based on the integration of many nearby orbits reaches 
a saturation limit when the spreading of the orbits extends to the whole domain 
of the phase space available to them. 

The main results of our investigation are: 

a) We justify theoretically why do 
metastable states with a constant rate of increase of the q-entropy appear 
when a system has weakly chaotic orbits. We find this to be due to the time 
evolution of the deviation vectors, which is of the form $\xi(t)\approx 
a t + e^{\lambda t}$, i.e. a combination of a linear and an exponential 
law, with $a>>1$ and $\lambda<<1$. This law in numerical applications appears 
as producing a transient behavior for a long time interval in which the growth 
of $\xi(t)$ is almost a power law $\xi(t)\propto t^p$, with $p>1$. Furthermore, 
the exponent $p$ can be associated to a $q-$ exponent via a simple relation 
$p=1/(1-q)$.

b) We introduce a method by which one can calculate the q-exponent along a flow 
of chaotic nearby orbits directly from the variational equations of motion.

c) The same method can be used as a `chaotic indicator' distinguishing weakly 
chaotic orbits from nearby regular orbits. We call this indicator APLE (average 
power-law exponent). Its sensitivity is comparable to that of other established 
indicators in the literature such as the Stretching Numbers \cite{Voglis01}, 
Fast Lyapunov Indicator \cite{Froes01}, Spectral Distance \cite{Voglis02}, 
the Mean Exponential Growth of Nearby Orbits \cite{Cincot01,Cincot03} or 
the Smaller Alignment Index \cite{Skokos01}.
In section 3 we make a comparison of the APLE with the FLI and the MEGNO.
While these indicators are practically equally powerful in distinguishing 
order from chaos, the APLE yields simultaneously the value of the q-exponent.

We give examples of the numerical behavior of APLE in the domain of weak 
chaos in 2D and 4D symplectic mappings. We consider in particular: 
a) a separatrix chaotic layer in the 2D standard map, b) a stickiness domain 
at the border of an island of stability in the same map, and c) weakly chaotic 
orbits in the Arnold web of resonances in a 4D symplectic map proposed in \cite{Froes03}.
In all these cases we identify which chaotic 
orbits exhibit the kind of metastable behavior proposed by Tsallis and his 
associates, and which fall into the usual regime of the constant Kolmogorov 
- Sinai entropy.  We also see how is this behavior depicted in the time 
evolution of the APLE, and of other indicators, i.e., the FLI or the 
MEGNO.

The definition of APLE is given in section 2, following a theoretical analysis of 
the  emergence of a transient power-law growth of deviations for weakly chaotic 
orbits in conservative systems. Section 3 presents numerical calculations in the 
2D standard map and in the 4D map proposed in \cite{Froes03}. Section 4 
summarizes the main conclusions of the present study.

\section{Theoretical considerations}

\subsection{Entropy and the growth of deviation vectors} 

\begin{figure}
      \centerline{\includegraphics[width=30pc] {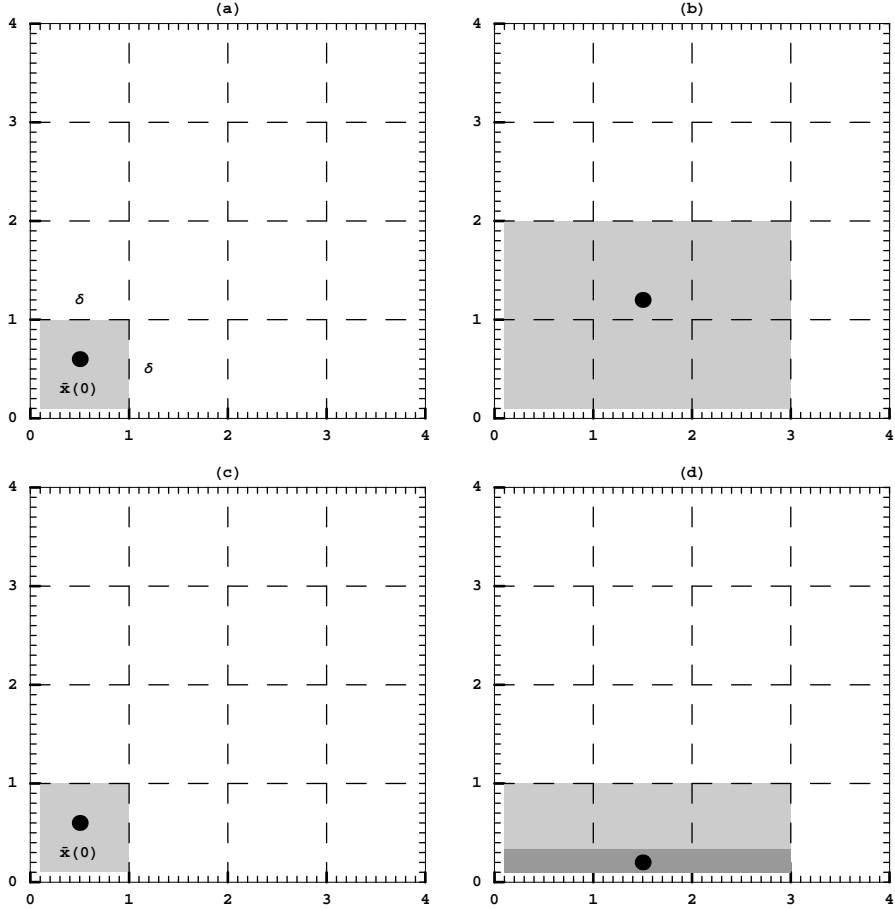}} 
      \caption{Schematic representation of the spreading of orbits in the 
neighborhood of a hyperbolic fixed point (origin) in the case of a 
non-conservative 2D dynamical system (a and b), or conservative system 
(c,d), see text for details. The numerical values in the coordinate axes 
are multiples of $\delta. $}
      \label{fig:Shannon}
\end{figure}

In order to introduce the relation of the concept of entropy to the solutions 
of the variational equations in dynamical systems, we consider the  
example of a system with a two-dimensional phase space (figure \ref{fig:Shannon}, 
schematic). In all the panels the origin O$\equiv(0,0)$ represents a hyperbolic fixed point. For simplicity, the eigenvectors of the monodromy matrix at O,  corresponding to a mapping of the orbital flow after a period of time $T$, 
are assumed orthogonal, while the eigenvalues $\lambda_1(T)$, $\lambda_2(T)$ 
are real and positive. The axes are parallel to the (orthogonal) eigenvectors.

We consider first a non-conservative case in which both $\lambda_1(T)$ and 
$\lambda_2(T)$ are greater than unity (figure \ref{fig:Shannon}a, in which we set 
$\lambda_1(T)=3$, $\lambda_2(T)=2$). Let the phase space in the neighborhood of 
O be divided into a number of square cells of linear size $\delta$. This size may represent 
an accuracy limit as regards our knowledge of the positions of the orbits due to 
experimental, numerical or other sources of uncertainty. The gray cell, of area 
$V_0=\delta^2$, is a set of initial conditions which are `nearby' to the initial 
condition $\bar{x}(0)$ inside the cell. At a time $T$, the orbit and the nearby 
orbits are mapped to the black point and the six gray cells shown in 
figure \ref{fig:Shannon}b respectively. If we ask whether the orbit is in the upper or 
lower, or, in the left or right group of cells, then after three questions of this type 
we can locate precisely in which cell the moving point is located, that is, we can 
have the maximum possible information on the position of the moving point in the 
phase space given the accuracy limit $\delta$. The number of questions 
needed to obtain this information is called Shannon's 
entropy (see \cite{Schuster01}). Asymptotically (in the limit of a large number of 
cells), Shannon's entropy is given by the logarithm of the number of gray cells 
$d=V(T)/V(0)$, where $V(T)$ is the total area (or `volume') of the gray cells 
at time $T$. This can be connected to the 
eigenvalues of the monodromy matrix at O via $V(T)=\lambda_1(T)\lambda_2(T)
\delta^2$, so that $S_{shan}=\ln d = \ln (V(T)/V(0)) = \ln\lambda_1(T)+
\ln\lambda_2(T)$. If the cells are interpreted as `micro-states', Shannon's 
entropy is equivalent to the usual Boltzmann-Gibbs entropy $S_{BG}=\ln d$. 
The average rate of increase of the Boltzmann-Gibbs entropy is then given by 
\begin{eqnarray}\label{sbgt1}
&{S_{BG}\over T}\equiv {S_{Shan}\over T} = \nonumber\\
&{1\over T}\ln\lambda_1(T)+{1\over T}\ln\lambda_2(T) 
= {1\over T}\big(\ln(\xi_1(T)/\xi_1(0))+\ln(\xi_2(T)/\xi_2(0))\big)
\end{eqnarray}
where $\xi_1(T)$,$\xi_2(T)$ are the time $T$ images of two initial 
deviation vectors $\xi_1(0)=(\delta,0)$, $\xi_2(0)=(0,\delta)$.  

If, now, we consider the case of a conservative system, in which $\lambda_1(T)$ 
and $\lambda_2(T)$ are reciprocal, the image of the set of orbits in the 
initial gray cell (figure \ref{fig:Shannon}c, in which we set 
$\lambda_1(T)=3$, $\lambda_2(T)=1/3$) 
is the dark gray area of the parallelogram shown in figure \ref{fig:Shannon}d. 
However, the vertical side of this parallelogram is now smaller than the accuracy 
limit $\delta$. Thus, if questions are made in order to identify in which cell 
the orbit is located, the search is restricted to a smaller number of 
cells $d=3$. In this case we have
\begin{equation}\label{sbgt2}
S_{BG}=\ln d = \ln\lambda_1(T) \Rightarrow
{S_{BG}\over T}={1\over T}\ln\lambda_1(T)={1\over T}\ln(\xi_1(T)/\xi_1(0))~~.
\end{equation}
In the limit of large $T$ the quantities $(1/T)\ln\lambda_k(T) = 
(1/T)\ln(\xi_k(T)/\xi_k(0))$, $k=1,2$ yield the spectrum of Lyapunov 
characteristic exponents of the orbits (in the neighborhood of $O$). We see 
that in both cases, of Eq.(\ref{sbgt1}) or Eq.(\ref{sbgt2}), the rate of 
increase of the Boltzmann - Gibbs entropy $S_{BG}/T$ is given by the sum of 
positive Lyapunov characteristic exponents of the orbits, which according to 
Pesin's theorem \cite{Pesin01}, is equal to the {\it Kolmogorov-Sinai} entropy of the 
orbital flow in the neighborhood of O. This means that the Kolmogorov-Sinai 
entropy is, in fact,  a measure of the rate of change of the entropy rather 
than a measure of the entropy itself.

The following is a more precise treatment of the previous schematic picture. 
Consider a partitioning of the n-dimensional phase space ${\cal M}$ of a 
coservative system into a large number of volume elements of size $\delta^n$ 
for some small $\delta>0$. 
Let $\vec{x}(0)$  be the initial condition of an orbit located in a 
particular volume element $V_0=\xi_{01} \xi_{02} ...\xi_{0n}$, where $\xi_{0k}$, 
$k=1,...,n$ are the linear dimensions of $V_0$ in a locally orthogonal set of 
coordinates in the neighborhood of $\vec{x}(0)$. Without loss of generality, 
we set all the initial values $\xi_{0k}$ equal, i.e., 
$\delta\equiv V_0^{1/n}=\xi_{0k}$, $\forall k$. All the orbits with 
initial conditions within $V_0$ are called `nearby' to the orbit 
$\vec{x}(t)$ with initial condition $\vec{x}(0)$. Because of the volume 
preservation, the orbital flow defines a mapping $V_0\rightarrow V(t)$ of the 
volume $V_0$ to an equal volume $V(t)$ at the time $t$. We want to find an 
estimate of the covering of the cells of ${\cal M}$ by the volume $V(t)$ in 
terms of the variational equations of motion. To this end, let 
\begin{equation}\label{dt}
\vec\xi(t)=D_t\vec\xi_0
\end{equation}
be the solution of the variational equations for an initial deviation vector 
$\vec\xi_0$ acted upon by a linear evolution operator $D_t$ determined solely 
by the orbit $x(t)$.
Let $\vec\xi_k'(t)$ be the images of $\vec\xi_{0k}$, $k=1,2,...,n$ under the 
action of the operator $D_t$. The vectors $\{\vec\xi_k'(t)\}$ form a complete 
basis of the tangent space to {\cal M} at the point $\vec{x}(t)$ 
iff $\{\vec\xi_{0k}\}$ form a complete basis of the tangent space to 
${\cal M}$ at the point $\vec{x}(0)$ and $rank(D_t)=n$. It is possible 
to obtain an orthogonal basis $\{\vec\xi_k(t)\}$ starting from $\{\vec\xi_k'(t)\}$ 
via the Gramm-Shmidt procedure \cite{Benet01}. The new basis is 
obtained by the recursive relation:
\begin{eqnarray}\label{grs}
\vec\xi_1(t) &= &\vec\xi_1'(t) \nonumber\\
\vec\xi_k(t) &= &\vec\xi_k'(t)
-\sum_{\nu=1}^{k-1}\big(\vec\xi_k'(t)\cdot\vec\xi_\nu(t)\big)
{\vec\xi_\nu(t)\over\xi_\nu(t)^2}~~.
\end{eqnarray}
The volume $V(t)$ is given by $V(t)=\xi_1(t)\xi_2(t) ...\xi_n(t)$. We 
reorder this basis by decreasing length of the vectors $\vec{\xi}_k(t)$. 
Let $V_c(t)$ be a coarse-grained volume equal to the total volume of all the 
cells visited by the orbits in $V(t)$. $V_c(t)$ is determined by only those 
vectors with lengths greater or equal to $\delta$, that is 
\begin{equation}\label{vc}
V_c(t) = \xi_1(t) \xi_2(t) ...\xi_m(t)\delta^{n-m}
\end{equation}
where $m$ is defined by the condition $m=\sup\{m': \xi_k(t)\geq \delta 
\mbox{ for all } k=1,\ldots,m'\}$. 
It follows that $m=n$ when $t=0$ and $V_c(0)=V(0)=\delta^n$. 

The Boltzmann - Gibbs entropy is defined by 
\begin{equation}\label{shabg}
	S_{BG}(t)\equiv \ln {W(t)}= \ln {\frac{V_c(t)}{V_c(0)}}
\end{equation}
where
\begin{equation}\label{shawdef}
W(t)={\xi_1(t)\xi_2(t)...\xi_m(t)\over\delta^m}
\end{equation}
yields the number of cells (or `micro-states') occupied by $V_c(t)$. 

The rate of growth of the entropy (\ref{shabg}) for a set of nearby orbits 
is related to the spectrum of the Lyapunov characteristic exponents of the 
reference orbit $\vec{x}(t)$. The Lyapunov characteristic exponents are 
given by
\begin{equation}\label{lce}
\lambda_k\equiv
\lim_{t\rightarrow \infty}{1\over t}\ln{\xi_k(t)\over\xi_k(0)}=
\lim_{t\rightarrow \infty}{1\over t}\ln{\xi_k(t)\over\delta}
\end{equation}
for $k=1,\ldots,n$, provided that the limits exist. 
Let $\lambda_1,\lambda_2,... \lambda_m$, $0\leq m\leq n/2$ 
be the set of positive exponents of the spectrum (\ref{lce}). According to 
Pesin's (1978) theorem, their sum is equal to the Kolmogorov-Sinai entropy 
$S_{KS}$ of the flow of orbits `nearby' to $x(t)$ \cite{Kolmog01,Sinai01}, i.e.
\begin{equation}\label{kslce}
	S_{KS}=\lambda_1+ \lambda_2+...+ \lambda_m~~~.
\end{equation}
The average rate of increase of the Boltzmann-Gibbs entropy up to the time 
$t$ is given by $S_{BG}(t)/t$. In view of Eq.(\ref{shawdef}), the limit of 
this rate, for $t\rightarrow \infty$ is
\begin{equation}\label{bgks}
	\lim_{t\rightarrow \infty} \frac{S_{BG}}{t}=\lim_{t\rightarrow
	\infty} \frac{1}{t} \ln {W} =\sum_{k=1}^m \lambda_k=S_{KS}
\end{equation}
Thus, in the limit $t\rightarrow \infty$ the Kolomogorov-Sinai entropy is 
equal to the asymptotic value of the mean growth rate of the Boltzmann-Gibbs 
entropy. Precise numerical examples of this relation were given in the case 
of low dimensional mappings \cite{Latora01,Latora02}. However, the 
orbits may exhibit transient or `metastable' states for long time intervals 
before reaching the limit (\ref{bgks}). Such states are characterized by 
a constant growth rate of the Tsallis q-entropy, to which we now turn our 
attention.

\subsection{Tsallis entropy and the Average Power Law Exponent}
The time evolution of the q-entropy \cite{Tsallis01} for an ensemble of orbits 
with initial conditions within the volume $V_0$ is given by
\begin{equation}\label{tse}
	S_q(t)=\frac{W(t)^{1-q}-1}{1-q}~~.
\end{equation}
In this equation $q$ is a constant parameter, known as the 'entropic index' $q$.
Dividing $S_q$ by $t/t_1$, where $t_1$ is a transient initial time of 
evolution of the orbits, and substituting $W$ from (\ref{shawdef}), yields 
the mean rate of evolution of $S_q$
\begin{equation}\label{tsdev}
	\frac{S_q}{t/t_1}=\frac{1}{(t/t_1)(1-q)}[ (\frac{\xi_1 \xi_2
	...\xi_m}{\delta^m} )^{1-q}-1]~~.
\end{equation}
For every $\xi_k$, $k=1,2,...m$, we define an {\it Average Power Law Exponent} 
(APLE) $p_k$ according to
\begin{equation}\label{devtim}
	\xi_k(t)=\xi_k(t_1)(\frac{t}{t_1})^{p_k},~~~~~k=1,2,...m~~.
\end{equation}
All the $p_k$ are, in general, functions of the time $t$, and the value of 
$p_k(t)$ yields the average logarithmic slope (or power-law exponent) of the 
evolution of $\xi(t)$ in the time interval from the time $t_1$ up to the time 
$t$. Furthermore, in conservative systems we have $p_1+p_2+...+p_m\geq 0$, 
since, by the preservation of volumes, the components $\xi_k(t)$ cannot be 
all decreasing functions of the time. 

In view of the definition of the APLEs (\ref{devtim}), equation (\ref{tsdev})) 
takes the form
\begin{equation}\label{tsti}
	{S_q\over t/t_1}={1\over (t/t_1)(1-q)}\big(
	\big({t\over t_1}\big)^{(p_1+p_2+...p_m)(1-q)}-1\big)~~.
\end{equation}
In the limit $t\rightarrow \infty$ the quantity $\frac{S_q}{t/t_1}$ tends to 
a non-zero finite value only if a) the $p_i$s take constant limiting values, 
and b) the entropic index $q$ satisfies the relation $(p_1+p_2+...p_m)
(1-q)=1$. In all other cases, $\frac{S_q}{t/t_1}$ tends either to zero or 
to infinity. If the deviations $\xi_k(t)$ grow asymptotically as a power law,  
then condition (a) is satisfied and the mean rate of increase of the Tsallis entropy 
$\frac{S_q}{t/t_1}$ tends to the sum of the positive  APLEs
\begin{equation}\label{tsle}
	\lim_{t\rightarrow \infty}\frac{S_q}{t/t_1}=p_1+p_2+...p_m
\end{equation}
for the value of $q$ given by
\begin{equation}\label{pqg}
	q=1-\frac{1}{p_1+p_2+...p_m}~~.
\end{equation}
In that case, if $p_1$ is by definition the maximum of all the APLEs, 
this exponent can be used as a lower bound of the limit of $\frac{S_q}{t/t_1}$, 
i.e.
\begin{equation}\label{tsmle}
	\lim_{t\rightarrow \infty}\frac{S_q}{t/t_1}\geq p_1~~.
\end{equation}
In practice we can use Eqs.(\ref{tsle}), (\ref{pqg}), or (\ref{tsmle}) for 
a long but finite time $t$ in order to estimate the average value of the 
q-exponent in the interval from $t_1$ and $t$, provided that this value is 
almost constant in this interval. Furthermore, the ratio of the length of 
the deviation vector $\xi^2(t) = \sum_{k=1}^m \xi_k^2(t)$ at $t$ with respect 
to the length of this vector at $t_1$ can be evaluated from the equation
\begin{equation}\label{xitbeta}
	\frac{\xi^2(t)}{\xi^2(t_1)}=\sum_{k=1}^m
	\frac{\xi_k^2(t)}{\xi^2(t_1)}=\sum_{k=1}^m
	\beta_k^2(\frac{t}{t_1})^{2p_k}
\end{equation}
where equation (\ref{devtim}) has been used and
\begin{equation}\label{betat}
	\beta_k^2=\frac{\xi_k^2(t_1)}{\xi^2(t_1)}
\end{equation}
with
\begin{equation}\label{betam}
	\sum_{k=1}^{m\leq n}\beta_k^2 \leq 1~~.
\end{equation}
Equation (\ref{xitbeta}) can also be written as
\begin{equation}\label{xitbeta2}
	\frac{\xi^2(t)}{\xi^2(t_1)}=(\frac{t}{t_1})^{2p_1}
        [\beta_1^2+\sum_{k=2}^m
	\beta_k^2(\frac{t}{t_1})^{-2(p_1-p_k)}]
\end{equation}
Since $p_1-p_k$ is positive for $2 \leq k \leq m$, the sum inside
the square brackets in the last expression tends asymptotically to
zero for $t>>t_1$. Thus we can write
\begin{equation}\label{aplebeta}
	APLE=p={\ln{\frac{\xi^2(t)}{\xi^2(t_1)}} \over
	2\ln{\frac{t}{t_1}}}=p_1 + \frac{\ln{[\beta_1^2+\sum_{k=2}^m
	\beta_k^2(\frac{t}{t_1})^{-2(p_1-p_k)}]}}{2\ln{\frac{t}{t_1}}}
\end{equation}
i.e in the limit of $t\rightarrow \infty $ APLE tends to $p_1$.

In two dimensional maps or in the Poincar\'{e} surface of section
of 2D Hamiltonian systems we have $m=1$, i.e. there is only one
positive exponent $p_1$ which is the limit of APLE p. In these
cases the APLE p is also the limit of the mean rate of growth of
Tsallis entropy $S_q$ with $q=1-1/p$ according to (\ref{pqg}).

\subsection{The time evolution of APLE for weakly chaotic orbits and the 
appearance of `metastable' states}

\begin{figure}
       \centerline{\includegraphics[width=30pc] {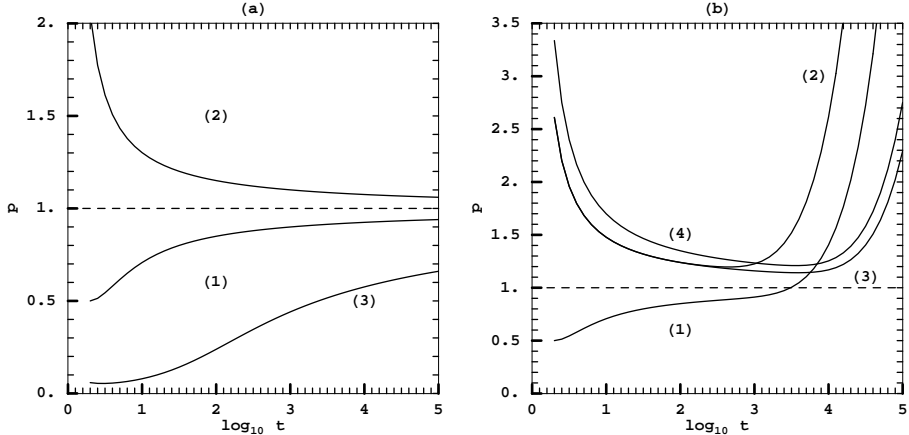}}
       \caption{ 
        The time evolution of $p$ vs. $ \log_{10} t $ for (a) regular orbits, 
        and (b) weakly chaotic orbits, according to the simplified model of 
        Eq.(\ref{soltoy}).
        In (a) we have (1) $ \xi = \sqrt{\Delta J_r^2+\Delta\phi^2} $ 
        and $ \xi_0=1 $, $ a = 0.5 $, $ \epsilon \rightarrow 0 $,  
       (2) $ \xi_0=1 $, $ a= 2 $, $ \epsilon \rightarrow 0 $, and (3) 
        $ \xi = 1+\sqrt{\Delta J_r^2+\Delta\phi^2} $
        with $\xi_0=10^{-2} $, $ a= 2 $, $\epsilon \rightarrow 0 $. 
        In (b) we have (1) $ \xi $ as in (a) with $\xi_0=1 $, $ a= 0.5 $, 
        $ \epsilon = 10^{-6} $, (the Lyapunov exponent and upper crossover 
        time are 
        $ \lambda \approx 10^{-3}, t_{2c} \approx 10^3$) 
        (2) $\xi_0=1 $, $ a= 3 $, $\epsilon =10^{-6} $, 
        ($ \lambda \approx 10^{-3}, t_{2c} \approx 10^3$)
        (3) $ \xi_0=1 $, $ a= 3 $, $ \epsilon = 10^{-8} $.
        ($ \lambda \approx  10^{-4}, t_{2c} \approx 10^4$)
        (4) $ \xi_0=1 $, $ a= 5 $, $ \epsilon = 10^{-8} $.
        ($ \lambda \approx 10^{-4}, t_{2c} \approx 10^4$)
        }
       \label{figthe}
\end{figure}

In the sequel we are interested in the behavior of the deviations $\vec{\xi}(t)$ 
for orbits in the border of a single resonance domain of a nonlinear Hamiltonian 
system of $n$ degrees of freedom. In the integrable approximation this border is 
separatrix-like. The Hamiltonian in resonant normal form (see e.g. \cite{Morbid01}) reads: 
\begin{equation}\label{hamac}
H = {J_{\psi}^2\over 2}-\omega_0^2\cos\psi+H_0'(J_2,\ldots,J_n)
\end{equation}
in action -angle variables $(J_\psi,J_2,\ldots,J_n)$ and $(\psi,\phi_2,
\ldots\phi_n)$. The resonant variables $(J_\psi,\psi)$, $J_\psi\in{\cal R}$, 
$\psi\in(\pi,\pi]$, obey a pendulum dynamics. 
The point $(J_\psi,\psi)=\pi$ defines a foliation of simply hyperbolic 
$n-1$ dimensional invariant tori of (\ref{hamac}) labelled by the constant 
values of $J_i$, $i=2,\ldots n$. In particular, when $n=1$ there is only a 
0-dimensional torus, i.e. an unstable equilibrium point, while if $n=2$ the 
tori are 1-dimensional, i.e., a family of unstable periodic orbits. 
The remaining phase space is foliated by n-dimensional tori. The deviations 
$\xi(t)$ on these tori  grow in general linearly $\xi(t)\approx \xi_0 + A~t$, 
where $A$ is a measure of the frequency differences between 
orbits on nearby tori. We may assume that the derivatives $|\partial\omega_i/
\partial J_j|\equiv |\partial^2 H_0'/\partial J_i\partial J_j|$ are bounded 
from above for all $i,j=2,\ldots,n$. This, however cannot be true for the 
frequency associated with either librations or rotations in the plane 
$(J_\psi,\psi)$, since the derivative of either the libration or rotation 
frequency with respect to the resonant action tends to infinity when the 
action tends to its limiting value on the separatrix, i.e.:
\begin{equation}\label{freqres}
\lim_{J_r\rightarrow J_{r,separatrix}}
\bigg|{\partial\omega_r(J_r)\over\partial J_r}\bigg|=\infty
\end{equation}
where, e.g., for librations 
\begin{equation}\label{jres}
J_r(E)=2\int_{\psi_{min}(E)}^{\psi_{max}(E)}J_\psi(E,\psi) d\psi
\end{equation}
is the libration action of an orbit labeled by the energy $E=J_\psi^2/2-
\omega_0^2\cos\psi$ and $\psi_{min}(E),\psi_{max}(E)$ are the limiting values 
of $\psi$ in the domain in which the latter equation has solutions.
Furthermore $\omega_r(J_r)=\partial E/\partial J_r$, and $J_{r,separatrix}=
J_r(E=\omega_0^2)$. A similar expression is found in the case of 
rotations, but with different limits of the integral (\ref{jres}). 
Near the separatrix, the value of $A$ in the linear term of the growth 
of deviations is determined essentially by the value of the derivative 
$a=|\partial\omega_r(J_r)/\partial J_r|$, which is finite, but large.
Thus the growth is essentially determined by the growth of the projection 
of the deviation vector on the plane $(J_r,\phi_r)$, where $\phi_r$ is the 
angle conjugate to $J_r$ according to the previous definitions. 
The variational equations for $\vec{\xi}\equiv(\Delta J_r,\Delta\phi_r)$ 
read
\begin{equation}\label{hamvar}
\dot{\Delta J_r} = 0,~~~\dot{\Delta\phi_r}={\partial\omega_r\over\partial J_r}
\Delta J_r
\end{equation}
yielding the solution $\Delta J_r(t)=C=$~const., $|\Delta\phi_r(t)|=
|\Delta\phi_r(0)+C~a~t|$. If $C=0$ the deviations remain constant 
$\xi(t)=|\Delta\phi_r(0)|$, while if $C\neq 0$ they grow linearly in time. 
In numerical applications we usually select a random orientation 
$\kappa = \Delta\phi_r(0)/\Delta J_r(0)\equiv\Delta\phi_r(0)/C$ for 
$\vec{\xi}_0$, so that, in general $C=\xi_0(1+\kappa^2)^{-1/2}$ is of 
order $C=O(\xi_0)\neq 0$ and $\xi(t)\approx \xi_0(1+at)$. The time behavior 
of the APLE for a deviation growing as $\xi(t)\approx \xi_0(1+at)$ is shown 
in figure \ref{figthe}a, when $a<1$ (curve (1)), or $a>1$ (curve (2)).
In fact, when $t$ is large and $C=O(\xi_0)$ we have $\xi(t)\approx 
\xi_0~a~t$, thus 
\begin{equation}\label{aplereg}
p\simeq {\ln(\xi(t)/\xi_0)\over \ln t} = {1\over \ln t}(\ln a +\ln t)~~.
\end{equation}
If $a<1$ we have $\ln a<0$ and $p$ tends to the value $p=1$ from below. 
On the other hand, if $a>1$ we have $\ln a>0$ and $p$ tends to the 
value $p=1$ from above. Thus, far from the separatrix the time evolution 
of $p$ is like in curve (1) of figure \ref{figthe}a, while close to the separatrix it is 
like in curve (2) of the same figure. Numerical examples of this behavior are 
given in section 3 below. Finally, independently of the distance from the 
separatrix, if the initial vector $\vec{\xi}_0$ is almost tangent to an 
invariant curve, we have $C<<\xi_0$. In that case we have $\xi(t)\simeq \xi_0$ 
and $p\approx 0$ for $t<1/C$, while for $t>1/C$ $p$ increases approaching 
asymptotically the value 1 from below, independently of the value of $a$ 
(figure \ref{figthe}a, curve 3). Numerically, we find that this behavior 
can only happen when the angle between $\vec{\xi}_0$ and the tangent to the 
invariant curve is small (below two degrees, see section 3).  Near this 
value there is a continuous transition in a narrow interval of values of 
$\phi$ from the curve (3) to the curve (2). Practically, when the initial 
orientation of $\vec{\xi}_0$ is selected randomly, in the great majority
of cases we encounter for regular orbits only the cases (1) and (2) of 
figure \ref{figthe}a. 

We now examine the time evolution of the APLE in the case in which a 
Hamiltonian perturbation is introduced, namely
\begin{eqnarray}\label{hamres}
H &= &{J_{\psi}^2\over 2}-\omega_0^2\cos\psi+H_0'(J_2,\ldots,J_n)
+\epsilon H_1(J,\phi;\epsilon) \nonumber \\
&=&H_r(J_r)+H_0'(J_2,\ldots,J_n)
+\epsilon H_1(J,\phi;\epsilon)~~. 
\end{eqnarray}
In Eq.(\ref{hamres}) we assume that an optimal resonant Birkhoff normal 
form has already been constructed (e.g. \cite{Morbid01}). This means that 
the action - angle variables in (\ref{hamres}) are obtained through a 
near-identity transformation from the original action - angle variables 
of the unperturbed Hamiltonian. Furthermore, in the Nekhoroshev regime 
the size of the perturbation $\epsilon$ in Eq.(\ref{hamres}) is 
exponentially small in the quantity $1/\omega_0^2$. Ignoring the small 
components of the deviation vector normal to the resonant plane $(J_r,\phi_r)$, 
the new variational equations of motion read:
\begin{eqnarray}\label{varper}
{d(\Delta J_r)\over dt} &\simeq 
&-\epsilon\big({\partial^2H_1\over\partial J_r\partial\phi_r}\Delta J_r
+{\partial^2H_1\over\partial\phi_r^2}\Delta\phi_r\big)  \nonumber\\
{d(\Delta\phi)\over dt} &\simeq 
&\epsilon\big({\partial^2H_1\over\partial J_r\partial\phi_r}\Delta\phi_r
+{\partial^2H_1\over\partial J_r^2}\Delta J_r\big) 
+ {\partial^2 H_r\over\partial^2J_r}\Delta J_r~~.
\end{eqnarray}

While a detailed exploration of the solutions of Eqs.(\ref{varper}) can 
only be made after $H_0$ and $H_1$ are known, we can explore the basic 
behavior of the solutions close to the separatrix limit by the following 
heuristic analysis. Close to the separatrix we have 
$|\partial^2H_r/\partial J_r^2| = |\partial\omega_r/\partial J_r|>>1$, 
while $\partial\omega_r/\partial J_r<0$. Assuming that all the partial 
derivatives of $H_1$ in (\ref{varper}) have O(1) average values over the 
basic periods of motion, we introduce the following simplified model yielding 
essentially the behavior of the resonant components of the deviation vector:
\begin{eqnarray}\label{vartoy}
{d(\Delta J_r)\over dt} &= 
&-\epsilon(\Delta J_r+\Delta\phi_r)  \nonumber\\
{d(\Delta\phi_r)\over dt} &= 
&\epsilon(\Delta J_r+\Delta\phi_r) -a\Delta J_r
\end{eqnarray}
for $a>1$ and $\epsilon<<1$. For $\epsilon=0$, Eqs.(\ref{vartoy}) take the  
form of the equations (\ref{hamvar}) of the integrable case. If we choose 
initial conditions perpendicular to the invariant curves of $H_r$, i.e., 
$\Delta J_r(0)=C=\xi_0$, $\Delta\phi_r(0)=0$, the solution of (\ref{vartoy}) 
reads:
\begin{eqnarray}\label{soltoy}
\Delta J_r(t) &= &\xi_0\bigg(
\cosh((a\epsilon)^{1/2}t)
-\big({\epsilon\over a}\big)^{1/2}\sinh((a\epsilon)^{1/2}t)
\bigg)
\nonumber\\
\Delta\phi_r(t) &= &\xi_0\bigg(
\big({\epsilon\over a}\big)^{1/2}-\big({a\over \epsilon}\big)^{1/2}
\bigg)
\sinh((a\epsilon)^{1/2}t)~~.
\end{eqnarray}
The asymptotic analysis of (\ref{soltoy}) yields now the reason why we 
observe a `metastable' behavior in the time evolution of deviations. 
If $\epsilon<<1$, then for times $t<(a\epsilon)^{-1/2}$ we can consider both 
$\epsilon$ and  $(a\epsilon)^{1/2}t$ as small quantities. Then, the first of 
equations (\ref{soltoy}) yields an almost constant term $\Delta J_r\approx \xi_0
+O(\epsilon t)$, while the second equation yields a linear behavior 
$|\Delta\phi_r| \approx \xi_0 |\epsilon -a| t$. This is similar to the case of
regular orbits. We thus have a behavior similar to the curve (2) of figure 
\ref{figthe}a. However, when $t>(a\epsilon)^{-1/2}$ the exponential behavior becomes 
dominant $\Delta J_r\sim\Delta\phi\sim\exp((a\epsilon)^{1/2}t)$, with Lyapunov 
exponent $\lambda=(a\epsilon)^{1/2}$. This behavior is 
exemplified in figure \ref{figthe}b, in which we plot $\xi(t)=\big(\Delta J_r^2(t)+
\Delta\phi_r^2(t)\big)^{1/2}$, with $\Delta J_r(t),\Delta\phi_r(t)$ given 
by Eq.(\ref{soltoy}), for different values of $a$ and $\epsilon$. We 
see that the combination of the linear and  exponential laws creates
a `plateau' of nearly constant value of $p$  between an initial time $t_{1c}\approx 10$ 
and a second time $t_{2c}$ which is essentialy given by $t_{2c}\approx \lambda^{-1}$.
These times are called `crossover times`. In the interval $t_{1c}\leq t\leq t_{2c}$ 
the deviation vector $\xi(t)$ grows almost as a power law $\xi(t)\propto t^p$ for 
a nearly constant value of $p>1$. We stress that the real time evolution 
$\xi(t) \propto a t + \exp(\lambda t)$ is mathematically different 
from a power law, and only a numerical resemblance to a q-exponential 
is systematically obtained for specific time intervals. As shown in figure 
\ref {figthe}b, the duration of the `metastable' behavior decreases when 
$\lambda$ increases, while the value at which $p$ is stabilized in the interval 
$t_{1c}<t<t_{2c}$ increases as $a$ increases. An analytical estimate of the 
plateau is obtained by noticing that the leading terms of Eq.(\ref{soltoy}) 
(for small and large times) yield the time evolution of $\xi(t)$ more precisely as 
$\xi(t)\approx 1+a t/2 + (a /2\lambda)\left(e^{\lambda t}-1\right)$. 
Taking $p\simeq\ln\xi/\ln t$ we readily find the profile of the function 
$p(t)$ in the neighborhood of a characteristic time $t_0=1/\lambda$, given 
by the following bounds for the first and second derivatives of $p$ (for $\lambda \ll 1$):
$$
{dp\over dt}\leq \lambda\bigg({2+1/e \over |\ln\lambda|}
+{\ln(\lambda+a e/2)\over(\ln\lambda)^2}\bigg)
$$
$$
{d^2 p\over dt^2} \simeq \lambda^2\left({1-2/e-1/e^2\over |\ln\lambda|}
+{1+\ln (a/2)-2/e\over (\ln\lambda)^2}+\frac{2 \ln(a e/2)}{|\ln \lambda|^3}\right)>0~~.
$$
The second of the above equations implies that the function $p(t)$ is convex 
at the time $t_0=1/\lambda$, so that the variations of $p$ over intervals 
$\Delta t$ around $t_0$ are bounded by the $O(\lambda/|\ln\lambda|)$ estimate 
for the first derivative, namely:
\begin{equation}\label{platan}
|\Delta p|\leq \lambda\bigg({2+1/e\over |\ln\lambda|}
+{\ln(\lambda+ae/2)\over(ln\lambda)^2}\bigg)\Delta t~~.
\end{equation}
For example, if $\lambda=10^{-6}$, a 1\% variation of the value of the APLE 
$\Delta p=0.01$ can only occur in an interval $\Delta t\sim 0.01|\ln/\lambda|
/\lambda$, or $\Delta t\sim 10^5$.  This value marks the extend of the plateau, which is 
a considerable fraction of the time $t_0\sim 1/\lambda$. The value of $p$ 
in this plateau is estimated as:
\begin{equation}\label{ppl}
p_{plateau}\simeq 1+{\ln(a e/2)\over\ln|\lambda|}
\end{equation}
The estimate (\ref{ppl}) is found to be in good agreement with the numerical 
values of $p$ given, e.g., in the examples below. Recalling that $a$ is essentially 
a measure of the derivative of the frequency $|\partial\omega_r/\partial J_r|$, we 
expect that, in a chaotic layer resembling a separatrix, but with some thickness, 
$p$ increases as we approach to the center of the layer, since $a$ increases abruptly, 
while $p$ is smaller near the edges of the layer. This is precisely what is found in 
numerical experiments, as analyzed in the following section. In particular, if one 
defines an average value $\overline{<p>}$ over the whole chaotic layer, 
this yields an average value of the entropic index $q=1-1/\overline{<p>}$ 
corresponding to the metastable behavior of the orbits in this chaotic layer. 

\section{Numerical applications of the APLE}

\subsection{Separatrix layer in the 2D Standard Map}
In the sequel we consider examples of numerical calculations of the 
APLE~$\equiv p$ in discrete conservative systems. The time $t$ in 
Eq.(\ref{aplebeta}) obtains discrete values $t=1,2,...$. In order to 
avoid a singular value of $p$ when $t=1$, we set $t_1=2$ and we 
calculate $p$ for $t\geq t_1$, with $p=0$ when $t=t_1$.  

\begin{figure}
       \centerline{\includegraphics[width=40pc] {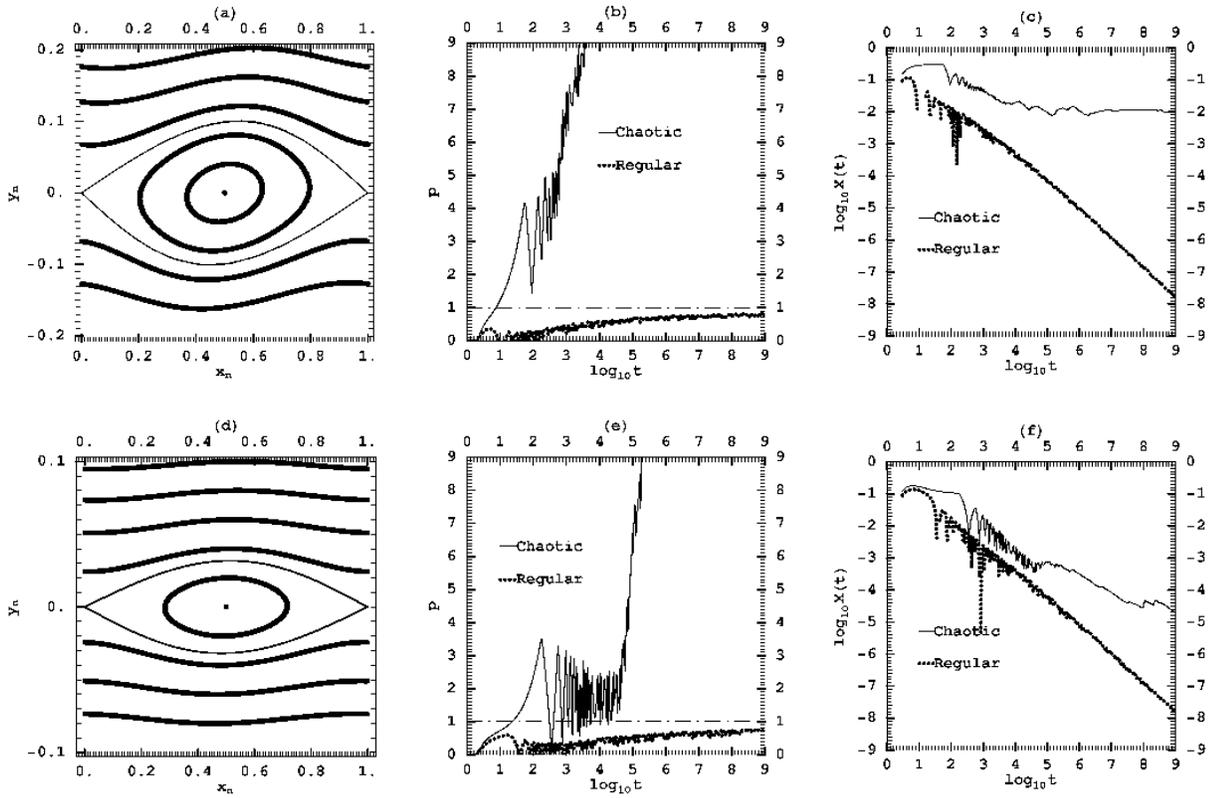}}
       \caption{
       (a) phase portrait of the standard map for $ K=10^{-1}$
       (b) The time evolution of APLE for a regular orbit in the libration region 
       of (a), and for a chaotic orbit in the thin separatrix layer, with initial 
       conditions along the unstable manifold of the periodic orbit $(0,0)$. 
       (c) the time evolution of the finite time Lyapunov number $\log_{10} \chi(t)$
       for the same orbits. (d,e,f) same as in (a,b,c), but for $K=10^{-2}$.
       }
       \label{fig:petsmall}
\end{figure}

We are particularly interested in the time evolution of $p$ for orbits 
near or within a domain of weak chaos. A basic example is provided by 
the 2D standard map \cite{Chirik01}:
\begin{eqnarray}\label{stmap}
        y_{n+1} &=& y_{n}+\frac{K}{2\pi}\sin{(2 \pi x_{n})}, \nonumber \\
        x_{n+1} &=& x_{n}+y_{n+1},
\end{eqnarray}
where $x_{n}$, $y_{n}$ are given modulo(1) in the intervals [0,1) and [-0.5,0.5) 
respectively. Figures \ref{fig:petsmall}a,d show the phase portrait of the map 
(\ref{stmap}) for $K=10^{-1}$ (figure \ref{fig:petsmall}a) and $K=10^{-2}$ 
(figure \ref{fig:petsmall}d). The thin solid lines in these figures are the 
unstable asymptotic manifolds emanating from the unstable periodic orbit 
$P_U\equiv(0,0)$. The manifolds are calculated by taking many initial conditions 
in a small segment of length $10^{-8}$ along the unstable eigendirection given 
by the monodromy matrix at $P_U$. Both phase portraits show the basic 
structure of the standard map for small $K$, i.e., a thin separatrix layer 
of weak chaos that separates a librational from a rotational domain. When $K$ 
is small the latter domains are filled almost entirely by invariant curves.

Figures \ref{fig:petsmall}b,e show the typical time evolution of APLE for regular 
or chaotic orbits. The dotted curves give the time evolution of the APLE  
for a regular orbit inside the domain of librations (initial conditions:  
$(x_0, y_0)=(0.0, 0.2)$, close to the stable periodic orbit at (0,0.5)), 
for $K=10^{-1}$ and $K=10^{-2}$. The initial deviation vector $\vec{\xi}(0)$ 
in both figures is chosen to be nearly perpendicular to the invariant curve 
passing through $(x_0, y_0)$. The temporal behavior of APLE in the dotted 
curves of figures.\ref{fig:petsmall}b,e is typical of regular orbits, i.e., the 
APLE grows slowly tending asymptotically to $p=1$ from below (as in curve (1) 
of figure \ref{figthe}a). The oscillations of $p$ around its local mean value 
are due to oscillatory variations of the component of the deviation vector 
$\vec{\xi}(t)$ locally tangent to the invariant curve. In fact, if we approximate 
the invariant curve by an ellipse, it can be shown that the amplitude of 
the oscillations of $p$ is proportional to the axial ratio of the ellipse. 
Furthermore, a plot of the finite time Lyapunov number
\begin{equation}\label{chit}
\chi(t)={1\over t}\ln|{\xi(t)\over \xi(0)}|
\end{equation} 
for the same orbits (figures \ref{fig:petsmall}c,f, dotted lines) shows also the 
behavior expected for regular orbits, i.e., $\chi(t)$ falls asymptotically 
as $t^{-1}$ for large $t$.

Now, the thin solid curves in figure \ref{fig:petsmall}b,e show the behavior of 
APLE for chaotic orbits inside the separatrix layers of figures 
\ref{fig:petsmall}a,d. In this case we take the initial conditions on the unstable 
manifolds of $P_U$, a fact ensuring that all the consequents of the chaotic orbits 
are on the same manifolds. The initial deviation vector $\vec{\xi} (0)$ is chosen 
perpendicular to the unstable manifold. We can immediately notice the difference 
in the time behavior of APLE for these two orbits. In the case of the solid 
curve of figure \ref{fig:petsmall}e ($K=10^{-2}$), the APLE grows initially crossing 
the value $p=1$ at a short crossover time $t_{1c}=25$. However, after this crossing 
the APLE describes a number of oscillations around a mean value that remains 
systematically above unity, up to a second crossover time $t_{2c}=38.000$. 
We find $<p>=1.9$ in the time interval $t_{1c} \leq t\leq t_{2c}$. As shown 
in figure \ref{fig:petsmall}f, the value at which the finite time Lyapunov number 
$\chi_(t)$ stabilizes is $LCN\equiv\lambda=10^{-4.8}$. We thus see that the 
crossover time $t_{2c}$ is essentially given by $t_{2c}\approx\lambda^{-1}$. 
On the contrary, in the case of the solid curve of figure \ref{fig:petsmall}b 
($K=10^{-1}$), the APLE grows from the start indefinitely, as expected for an 
exponential growth of deviation vectors, and there is no visible `metastable' 
behavior in the time evolution of $p$. In that case the Lyapunov number 
(figure \ref{fig:petsmall}c) is rather large $\lambda = 10^{-2}$, and the 
corresponding crossover time $t_{2c}\approx 10^2$ is of the same order as 
$t_{1c}$, i.e., extremely short to produce any visible effect.

As in the case of regular orbits, the oscillations of APLE around a local mean 
value in figures \ref{fig:petsmall}b,e are due to the oscillatory behavior of the 
component of the deviation vector $\vec{\xi}(t)$ which is tangent to a 
theoretical separatrix passing through the center of the separatrix chaotic 
layer. In this case we find that the first maximum value of $p$ in 
figures \ref{fig:petsmall}b,e occurs when the orbits pass close to the first 
homoclinic point of the unstable manifold emanating from $P_U$ and the 
stable manifold emanating from the image of $P_U$ modulo 1.

In order to check the dependence of the time evolution of the APLE on the 
initial orientation of the deviation vector $\vec{\xi}(0)$, corresponding 
to the value of the constant $C$ in the theoretical analysis of subsection 
(2.3), the following numerical test is performed: Starting from an arbitrary 
initial orientation of the deviation vector $\vec{\xi}(0)$, the orbit and 
the variational equations are integrated for a long time $t=\tau$. It is well 
known (see, for example \cite{Voglis03}) that the deviation vector 
$\vec{\xi}(t)$ evolves so that it tends to become tangent to the invariant 
curve, in the case of a regular orbit, or parallel to the direction of a 
nearby unstable asymptotic curve (invariant manifold), in the case of a 
chaotic orbit. Let $(x_{\tau}, y_{\tau})$ be the position of the orbit at 
$t=\tau$ and $(dx_{\tau},dy_{\tau})$ be the components of the deviation 
vector at this time. Taking $\tau$ long enough so as to ensure that the 
limit of tangency was reached down to the numerical precision level, 
we use $(x_{\tau}, y_{\tau})$ and  $(dx_{\tau}, dy_{\tau})$ as the initial 
conditions of a second orbit and of its variational equations. We then compare 
the time evolution of the APLE for this deviation vector and for a second 
deviation vector associated to the same orbit, but with orientation forming 
an angle $\phi$ with $(dx_{\tau}, dy_{\tau})$.
\begin{figure}
       \centerline{\includegraphics[width=40pc] {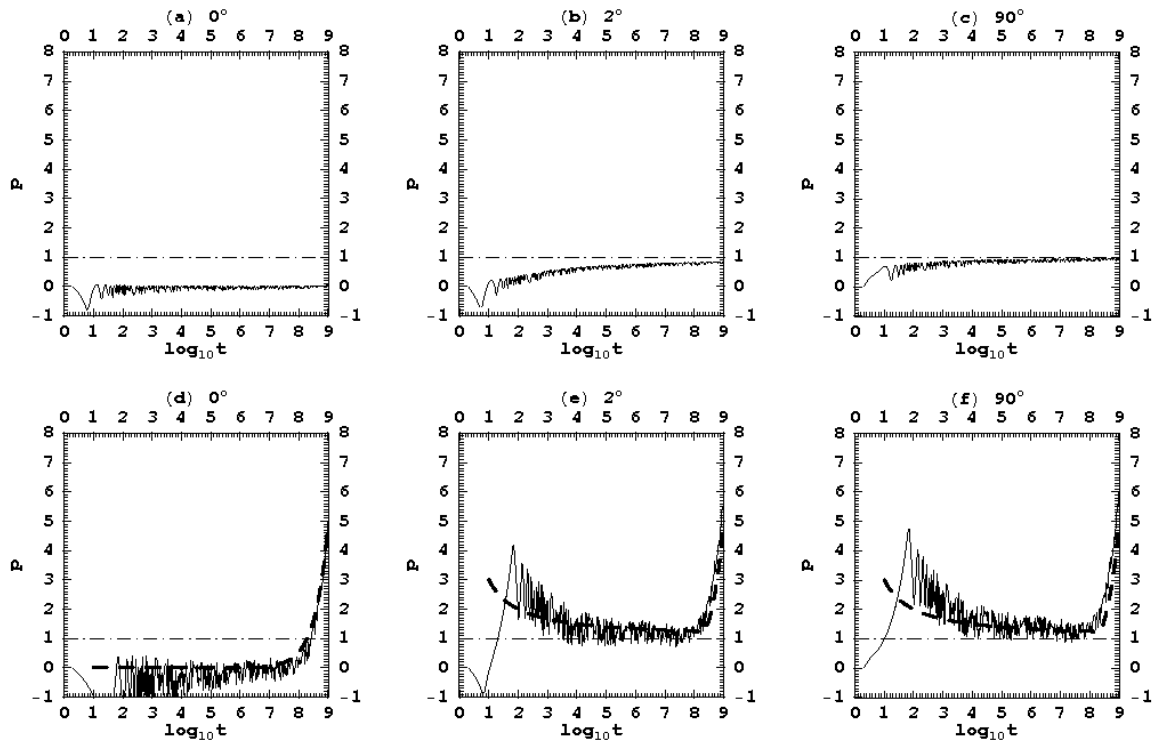}} 
       \caption{
       The time evolution of $p$ along a regular orbit with initial conditions 
       $(x_0,y_0) =$ $(0.5000046134885427,$$0.08097827536667734)$,
       for $ K=0.1 $ and three different deviation vectors forming an initial 
       angle (a) $0^0$, (b) $2^0$, (c) $90^0$ with the tangent to the invariant curve. 
       The corresponding evolution for a chaotic orbit (initial condition 
       $(0.5029208124108719,$$0.09994913824229859)$ is shown in (d-f), for $ K=0.1 $. 
       The bold dashed line in (d) is the theoretical solution corresponding to an 
       exponential law $\xi(t)/\xi_0=e^{\lambda t}$ with $\lambda=10^{-6.8} $, 
       while in (e),(f) we superimpose to this a linear term $a t$ with $a=2$.}
       \label{fig:angdep}
\end{figure}


Figure \ref{fig:angdep} shows examples of the dependence of APLE on the initial 
orientation using three
different orientations, namely $\phi\simeq 0$, $\phi=2^o$ and $\phi=90^o$, 
in the case of a regular orbit with $K=0.1$ (first row in figure \ref{fig:angdep}),  
and in the case of a chaotic orbit with $K=0.1$ (second row in figure \ref{fig:angdep}).  
Clearly, the evolution of APLE is sensitive on $\phi$ only for small 
values of $\phi$. For example, in the case of the regular orbit, when 
$\phi\simeq 0$ (figure \ref{fig:angdep}a) we have $\Delta J=C\simeq 0$ and 
the corresponding term in the solution of Eq.(\ref{hamvar}) is suppressed. 
Thus $\xi(t)$ simply makes oscillations around the mean value $<\xi(t)>=
\xi(0)$, and the APLE remains close to a zero value even after $t=\tau=10^9$ 
iterations. On the other hand, when $\phi$ is equal to only $\phi=2^o$, 
the time evolution of APLE becomes already very similar to its typical 
behavior, as concluded by a comparison with the case $\phi=90^o$, 
corresponding to an initial deviation vector perpendicular to the invariant 
curve. The same phenomena apply to the case of weakly chaotic orbits as 
in the second row of figure \ref{fig:angdep}. In that case, the asymptotic 
limit for all three values of the initial angle $\phi$ is an exponential 
growth of the corresponding deviation vectors, leading to a constant limit 
of the Lyapunov number $\lambda \simeq 10^{-6.8}$. When $\phi=0$ the 
`metastable' behavior does not show up in the time evolution of the APLE. 
However, this behavior is clearly seen when $\phi\geq 2^0$, and it lasts 
up to the crossover time $t_{2c} \sim 10^7>10^{6.8}=1/\lambda$. In fact, plotting 
the theoretical solution (\ref{soltoy}) corresponding to a choice 
$\phi=90^0$ in this case shows a good agreement with the numerical 
results for both $\phi=2^0$ or $\phi=90^0$. 

\subsection{The APLE as a chaotic indicator measuring the entropic q-index. 
Comparison with FLI and MEGNO}

In theory, the use of APLE as a `chaotic indicator' distinguishing regular from 
chaotic orbits is straightforward. In the case of regular orbits the value of 
APLE tends to $p=1$ as $t\rightarrow\infty$, while in the case of chaotic orbits 
we have $p\rightarrow\infty$ as $t\rightarrow\infty$. 
In practice, however, one can only evaluate $p$ over a finite integration time 
$T$ and a numerical indicator of chaos is considered as efficient if this time 
is small. An example of efficient chaotic indicator that is widely used in the 
literature is the Fast Lyapunov Indicator (FLI) \cite{Froes01}
in its revised form \cite{Froes02}. Let $\xi(t)$ be the length of 
the deviation vector of an orbit. The revised form of FLI reads:
\begin{equation}\label{fli}
FLI = \sup\{\ln\big(\xi(t)/\xi(0)\big), 0\leq t\leq T\}~~.
\end{equation}
In the case of regular orbits the deviations grow linearly, so that 
$\ln\big(\xi(t)/\xi(0)\big)\approx\ln t$. We can thus set a threshold value, 
say $FLI_0(T)=\ln 10+ \ln T$, corresponding to a deviation vector larger 
by a factor ten from the one corresponding to the linear growth of deviations 
up to the time $t=T$. 
Then, if $FLI>FLI_0$ the orbit is called chaotic, otherwise it is called regular. 
In fact, if we are close to the border of a separatrix chaotic layer, 
we have $\xi(t)=\xi_0(1+a~t)$ for regular orbits, with $a>>1$ (subsection 
2.3). Thus $\ln\big(\xi(t)/\xi(0)\big)\simeq \ln a+\ln t$. If we choose the 
threshold of $FLI_0$ as above, then, if $a>10$, a regular orbit can be 
erroneously characterized as chaotic. Thus, an improved formula for the 
threshold value is 
$$
FLI_0(T)=\ln 10 + \ln a  + \ln T
$$ 
where the value of $a$ can be estimated by the value of $\xi(t_{1c})$ for a time 
$1<<t_{1c}<<T$, since, if $a>>1$, we have $\xi(t_{1c})\simeq \xi_0 a t_{1c}$. Then, the 
orbit is considered as chaotic if $FLI>FLI_0$ at the time $T$. 

\begin{figure}
       \centerline{\includegraphics[width=30pc] {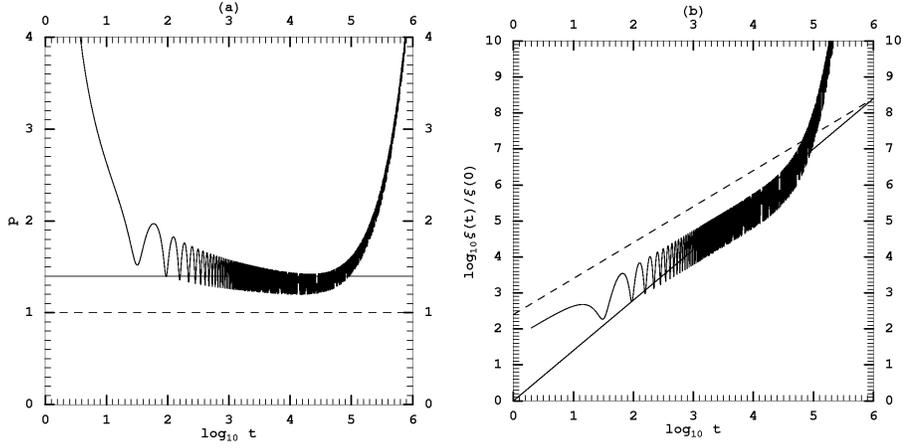}} 
       \caption{
       (a) A theoretical example of evolution of the APLE when 
       an oscillatory term is added to the solution (\ref{soltoy}), namely 
       $\xi = \sqrt{\Delta J_r^2+\Delta\phi^2}(1+\cos{0.1 t}) $ with 
       $\Delta J_r$, $\Delta\phi_r$ given by Eq.(\ref{soltoy}) and $\xi_0=1$, 
       $a=30$, $\epsilon=10^{-10}$. (b) The evolution $ \log_{10}
       (\frac{\xi (t)}{\xi(0)})$ vs. $\log_{10}t$ for the same example. The 
       dashed line corresponds to a linear law $\ln\xi(t) =\ln a +\ln t+\ln 10 $, 
       while the continuous line indicates a power-law $\xi(t)\propto t^p$ with 
       ($p=1.4$, i.e., as given by the approximate plateau of (a).
        }
       \label{fig:flicha4}
\end{figure}
\begin{figure}
       \centerline{\includegraphics[width=40pc] {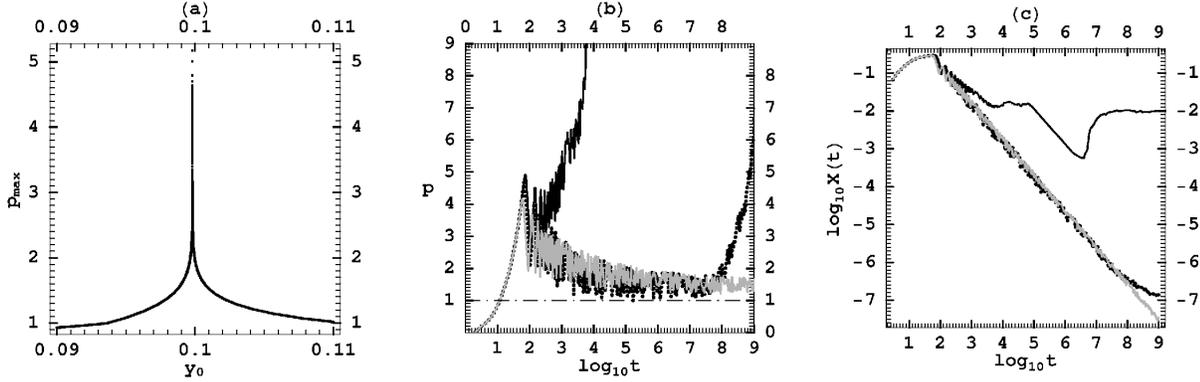}} 
       \caption{
       (a) The maximum value $ p_{max}$ in a time interval $ 2 < t \leq 10^3$ 
        as a function        of the initial condition $y_0$ of the orbits in a 
       segment along the line $x_0 = 0.5 $ 
       crossing the separatrix domain of figure \ref{fig:petsmall}a. 
       The peak value is for the orbit with initial conditions $(x_0,y_0)=
       (0.5,0.0998071243)$, and initial deviation vector $\vec{\xi_0}= 
       (\frac{1}{\sqrt{2}},\frac{1}{\sqrt{2}})$. The evolution of the APLE and of 
       $ \log_{10}\chi(t)$ for the same orbit are shown as solid black curves in 
       (b) and (c) respectively. The dotted black and the gray curves in the same 
       plots correspond to orbits with initial conditions $(0.5,0.099807124)$ and 
       $(0.5,0.09980712)$ respectively.
       }
       \label{fig:gridsm}
\end{figure}

Suppose now that the orbit is weakly chaotic, so that a `metastable' behavior 
exists for which $\xi(t)\sim \xi(t_{1c})(t/t_{1c})^p$, with $p>1$ (figure \ref{fig:flicha4}a, in 
which a plateau is formed at about $p=1.4$, from $t_{1c}=10^3$ to $t_{2c}=10^5$). 
As shown in figure \ref{fig:flicha4}b, the value of $\ln\big(\xi(t)/\xi(t_{1c})\big)$ crosses 
the line of $FLI_0(t)$ (solid straight line) at about the same time ($T=10^5$) 
when the slope of the quantity $\ln\big(\xi(t)/\xi(0)\big)$ vs. $\ln t$ 
crosses the value $p=1.4$ upwards, i.e. $t_{2c}\approx T$. This means that 
the APLE yields the characterization of the orbit as chaotic at the time 
$t_{2c}$ which is of the same order as the minimum time $T$ needed by the FLI. 
In fact, after $t>T$ the exponential growth of deviations 
becomes dominant and the slope of $\ln\big(\xi(t)/\xi(0)\big)$ vs. $\ln t$ 
tends very quickly to infinity. 

In practice, we found that the location of thin chaotic layers in resonances can 
be determined using APLE in a way analogous to the Eq. (\ref{fli}), 
namely:
\begin{equation}\label{pmax}
p_{max}=\sup\big\{p={\ln(\xi(t)/\xi(t_1))\over\ln(t/t_1)},t_1 < t\leq T\big\}~~.
\end{equation}
In order to probe numerically the sensitivity of $p_{max}$ to thin chaotic 
layers, figure \ref{fig:gridsm}a shows the variation of the value of $ p_{max}$, 
for $ 2 < t \leq T=10^3$, along a segment of the line of initial conditions 
$x_0 = 0.5$ passing through the separatrix chaotic layer of figure 
\ref{fig:petsmall}a. As the center of the resonance is approached, the value 
of $p_{max}$ increases abruptly, the peak value marking clearly the center of 
the chaotic layer. The peak value of $p$ corresponds to an orbit with initial 
conditions $(x_0,y_0)\equiv$ $(0.5,0.0998071243)$ (the initial deviation 
vector is taken as $\vec{\xi}_0 = (\frac{1}{\sqrt{2}},\frac{1}{\sqrt{2}})$). 
Since the chaotic layer is very thin, the behavior of APLE in a thin 
domain including this orbit is very sensitive on the choice of initial 
conditions. Thus, the central orbit has Lyapunov number $LCN\simeq 10^{-2}$ 
(figure \ref{fig:gridsm}c), and it shows only a small plateau in the values of 
$p$ at $p\simeq 3.5$, for about 900 periods, i.e., from $t=10^2$ to $t=10^3$ 
(figure \ref{fig:gridsm}b). However, if we only change the initial value $y_0$ by 
cutting the last digit, $(x_0,y_0)\equiv$ $(0.5,0.099807124)$, the new orbit has 
a much smaller Lyapunov number ($LCN\simeq 10^{-7}$, and the APLE forms a 
conspicuous plateau at $p\simeq 1.5$ which lasts for $10^7$ periods. Finally, 
if we cut one more digit in $y_0$, $(x_0,y_0)\equiv$ $(0.5,0.09980712)$, 
the orbit shows no sign of chaos up to $t=10^9$, although the plateau formed 
in the values of APLE (gray curve in figure \ref{fig:gridsm}b) is definitely 
at values $p>1$, an indication that the orbit may finally be chaotic. We notice 
that these results depend also on the machine precision, and different runs 
with different machines or precision levels yield qualitatively the same picture 
as in figure \ref{fig:gridsm}, but for somewhat different choice of initial 
conditions. 
 
\begin{figure}
      \centerline{\includegraphics[width=40pc] {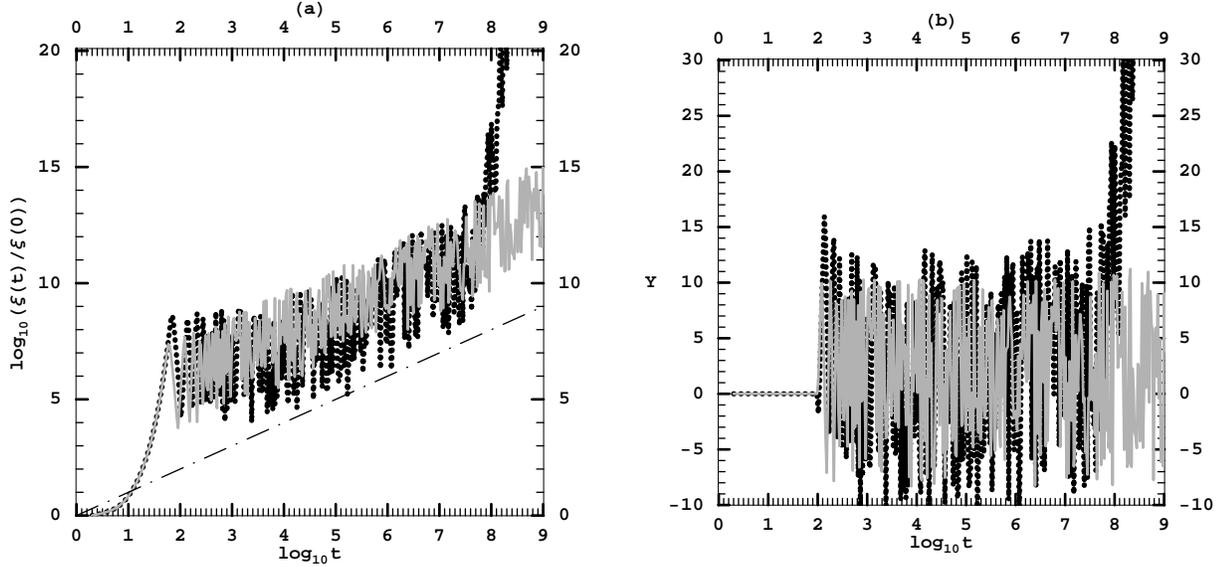}} 
      \caption{ 
      (a) The time evolution of $\log_{10}(\frac{\xi (t)}{\xi(0)})$, and 
      (b) the time evolution of the MEGNO ($Y$) for the same orbits (gray 
          and black dotted) as in figure \ref{fig:gridsm}b,c.}
      \label{fig:flimeg}
\end{figure}
\begin{figure}
	\centerline{\includegraphics[width=20pc] {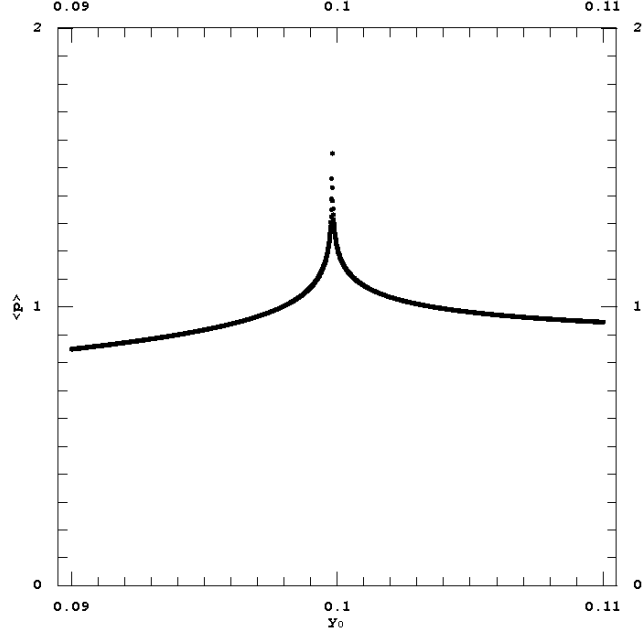}} 
        \caption{
        As in figure \ref{fig:gridsm} but for the average value $<p>$ 
        calculated over an interval of $ 10^4$ iterations per orbit. 
         }
	\label{fig:averpsm}
\end{figure}

Figure \ref{fig:flimeg}a shows, for comparison, the identification of the same 
orbits as in figure \ref{fig:gridsm} by the FLI. Clearly, the two methods have 
a similar sensitivity, i.e., $T\simeq t_{2c}=10^8$ for the dotted black curve 
of figure~\ref{fig:flimeg}a. Figure \ref{fig:flimeg}b shows the behavior, for the 
same orbits, of yet a different chaotic indicator: the MEGNO \cite{Cincot01}. 
The definition of MEGNO for continuous systems is:
\begin{equation}\label{megno}
Y(T)={2\over T}\int_{0}^T{\dot{\xi}(t)\over\xi(t)}tdt~~
\end{equation}
whereas, in the case of mappings, $\dot{\xi}(t)$ is replaced by the finite 
difference of the deviations $\xi$ at successive time steps. If the deviations 
have an average power law behavior $\xi(t)\sim t^p$ in the interval 
$0\leq t\leq T$, Eq.(\ref{megno}) yields the value of MEGNO $Y=2p$. 
However, this is not so when the power-law is transient and approximate, 
i.e., it comes from a combination $\xi(t)\approx a~t + e^{\lambda t}$. 
In fact, for regular orbits we readily find $Y(T)=2(1-\ln(1+aT)/(aT)$, 
thus $Y(T)<2$ for $T>1/a$ even if $a>>1$, that is the MEGNO cannot 
yield a power-law exponent $p>1$. Furthermore, we find that the numerical 
behavior of the MEGNO shows variations which are about one order of 
magnitude larger than those of APLE in the whole interval of `metastable' 
behavior of the orbits. This is 
exemplified in figure \ref{fig:flimeg}b, referring to the same orbits 
as in figure \ref{fig:gridsm} or \ref{fig:flimeg}a. The fast oscillatory 
variation of the MEGNO, of amplitude $\Delta Y \approx 10 $, shown in figure 
\ref{fig:flimeg}b results in that, while the maximum value of $Y$ is positive 
and above the threshold $Y=2$, the minimum value is negative. The mean 
value $<Y>$ in the transient interval of time is below $2$. 
On the other hand, not only the corresponding values of the 
APLE (figure \ref{fig:gridsm}b) are always positive, but they are clearly 
above the threshold $p=1$ during the whole interval $0\leq t\leq T$. Thus, 
the identification of the metastable behavior is much more clear by 
the APLE than by the MEGNO, and the APLE can be used to obtain a useful 
quantitative estimate of $<p>$, and thus of the entropic index $q=1-1/<p>$. 
This is shown in figure (\ref{fig:averpsm}), which yields the average value 
of $<p>$ along an orbit in the time interval $0\leq t\leq T=10^4$, as a 
function of the orbit's initial condition, for all the orbits in the same 
scanning of a thin chaotic layer as in figure \ref{fig:gridsm}. The general 
structure of this diagram is as in figure \ref{fig:gridsm}a. However, only 
a subset of these orbits have $<p>$ larger than unity. These orbits yield 
an average value of the values of $<p>$ equal to $\overline{<p>}=1.07$, 
corresponding to an entropic index $\bar{q}=0.065$.  

\subsection{Stickiness region}
A case of particular interest, in which weak chaos emerges, is in the stickiness 
region at the border of an island of stability. The stickiness of the orbits 
is due to the existence of one or more cantori (see \cite{Meiss01} for a review). 
The cantori arise from the destruction of KAM tori, which, according 
to Greene's criterion \cite{Greene01} happens when the stable periodic orbits 
with rotation numbers forming sequences with limit equal to the rotation 
number of a torus become unstable \cite{Cont01,Cont03,Efthym01,Efthym02}.
Since the dynamics at this limit is very close to hyperbolic, we expect 
that weakly chaotic orbits in the stickiness domain exhibit a transient 
metastable behavior for times of the order of the stickiness time. 

\begin{figure}
	\centerline{\includegraphics[width=40pc] {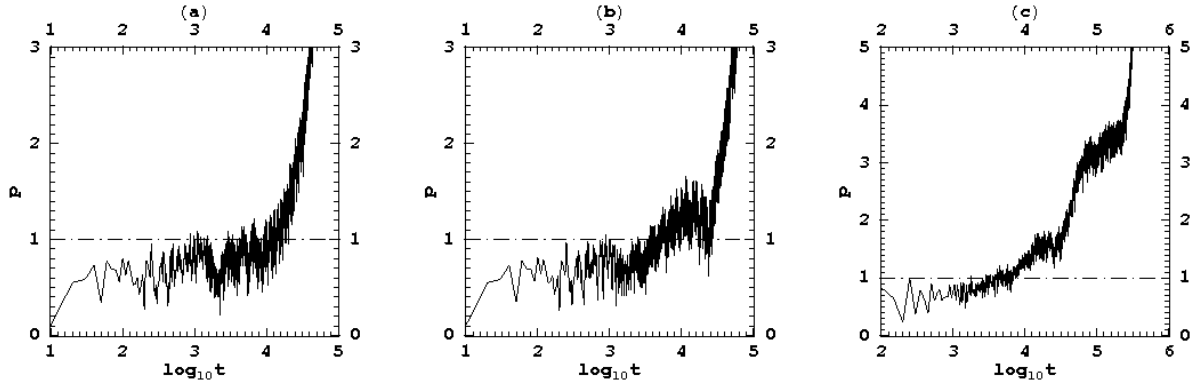}} 
        \caption{
         The time evolution of the APLE for some orbits belonging 
         to the sticky region of an island of stability in the standard map for 
         $ K=5$ 
         (see \cite{Cont02}). The initial conditions of the orbits are 
         (a) $(x_0,y_0)=$ $(0.64337,0.36)$, (b) $(x_0,y_0)=$ $(0.6433684,0.36)$ 
         and (c) $(x_0,y_0)=$ $(0.643363,0.36)$. 
        }
	\label{fig:stick}
\end{figure}
In order to study this phenomenon, we 
take initial conditions at the border of an island of stability as in 
\cite{Cont02}, namely we consider initial conditions 
in a line segment with $y_0=0.36$ and $0.64324\leq x_0\leq 0.6434$ 
(as in figure 7 of \cite{Cont02}). The stickiness time is 
particularly high (it can be larger than $10^6$ periods) when $x_0>0.64336$. 
By integrating many chaotic orbits in this domain, we found three kinds 
of different behavior in the time evolution of the APLE, shown in 
figures (\ref{fig:stick})a,b,c respectively: (a) An orbit may form no 
`plateau' beyond $p=1$ until the orbit escapes in the large chaotic 
sea outside the stickiness zone. (b) An orbit may form one plateau 
lasting for times of the order of its stickiness time, e.g. the orbit 
of figure \ref{fig:stick}b has a clear metastable behavior, with $p\simeq 
1.3$ for $7000\leq t\leq 25000$, while the orbit escapes after about 
$10^5$ periods. (c) An orbit forms more than one `plateaus' before 
escaping (the orbit of figure \ref{fig:stick}c forms two large plateaus 
at $p\simeq 1.5$ and $p\simeq 3$, lasting for about $10^4$ and $10^5$ 
periods respectively, while the stickiness time is about $5\times 10^5$ 
periods). The kind of behavior encountered by these orbits is very 
sensitive to the choice of initial conditions, a fact consistent 
with the fractal structure of the phase space near cantori. 

\subsection{The Arnold Web of a 4D Map}

Chaotic indicators are widely used in order to visualize the Arnold web 
of resonances in multidimensional conservative systems. In the following 
numerical examples we consider the 4D symplectic mapping proposed in 
\cite{Froes03}
\begin{eqnarray}\label{frm}
   x_{j+1} & = & x_j-\epsilon \frac {\sin{(x_j+y_j)}}
   {(\cos{(x_j+y_j)}+cos{(z_j+t_j)}+4)^2} \nonumber \\
   y_{j+1} & = & y_j+x_j 
   ~~~~~~~~~~~~~~~~~~~~~~~~~~~~~~~~~~~~~~~~~~~~~~~~~~~~~ (mod 2 \pi) \nonumber \\
   z_{j+1} & = & z_j-\epsilon \frac {\sin{(z_j+t_j)}}
   {(\cos{(x_j+y_j)}+cos{(z_j+t_j)}+4)^2}  \\
   t_{j+1} & = & t_j+z_j 
   ~~~~~~~~~~~~~~~~~~~~~~~~~~~~~~~~~~~~~~~~~~~~~~~~~~~~~ (mod 2 \pi) \nonumber
\end{eqnarray}
in order to study the behavior of APLE at the chaotic border of a single 
resonance domain. In this mapping, $(x_j,z_j)$ are action variables and 
$(y_j,t_j)$ are their conjugate angles. When $\epsilon=0$ 
we have constant values of the actions $x_{j+1}=x_j=x$, $z_{j+1}=z_j=z$, 
while the values $x,z$ yield also the frequencies, i.e., the per step changes 
of the angles. The lines $m_1x+m_2z$, with $m_1,m_2$ integer, 
yield the Arnold web of resonances in the action plane. The chaotic motions 
in this mapping along the borders of resonances, when $\epsilon\neq 0$, were 
studied in detail in \cite{Guzzo01,Froes03}. Of particular interest are the
chaotic motions when $\epsilon$ is smaller than a threshold value 
$\epsilon<\epsilon_0$ marking the onset of validity of the Nekhoroshev regime 
\cite{Nekh01} (see \cite{Guzzo01}). In that case 
the chaotic motions can be identified to the phenomenon of `Arnold diffusion' 
\cite{Arnold01} (see \cite{Cincot02} for a review). 

\begin{figure}
        \centerline{\includegraphics[width=40pc] {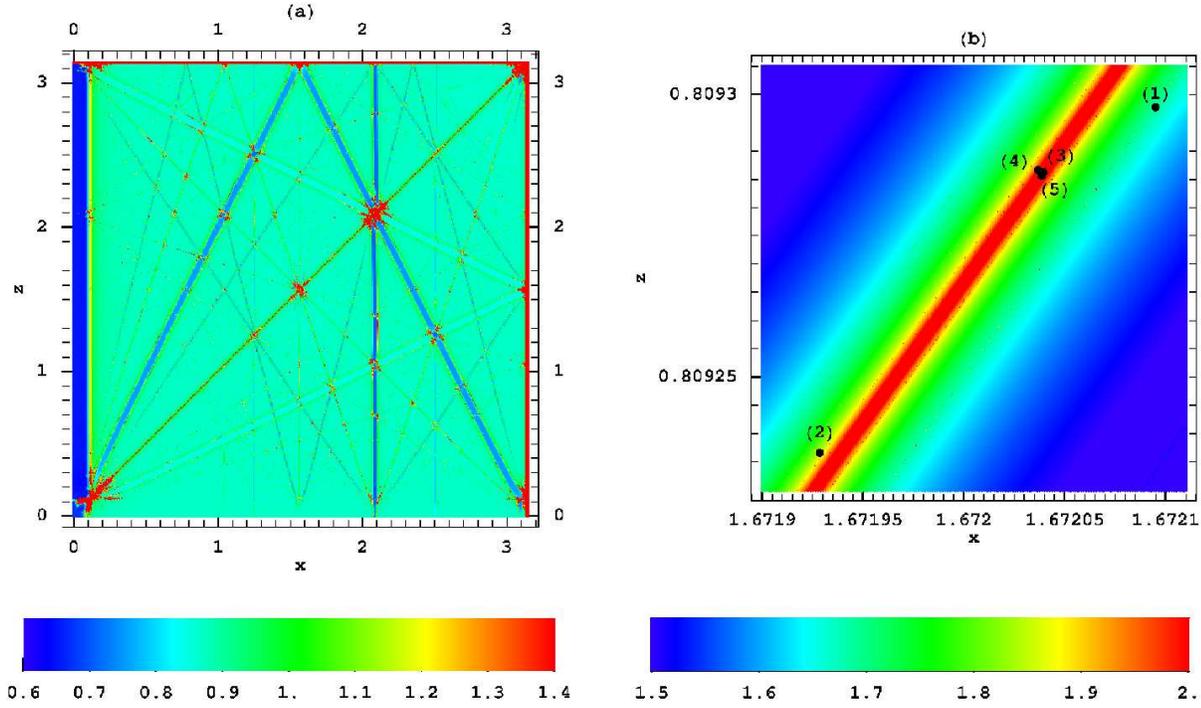}} 
        \caption{
         (a) A plot of the geography of the resonances in the action 
         space of the 4D map (\ref{frm}) when $\epsilon =0.05$, by means of the 
         index $p_{max} $ calculated over $T=10^4$ iterations per orbit. The 
         initial conditions are in a $ 500 \time 500 $ uniform grid and 
         the initial deviation vector for each orbit is 
         $ \vec{\xi} = (\frac{1}{\sqrt{2}},0,0,\frac{1}{\sqrt{2}})$.  
         Each point is colored according to the value of $p_{max}$ in the range 
         $0.6 \leq  p_{max} \leq 1.4$.
        (b) A zoom of (a) around the resonance $\frac{x_j}{z_j}=2 $,
        produced from a set of $ 500 \times 500 $ initial conditions and the same vector 
        $\vec{\xi}$ as in (a). The color scale for $p_{max}$ is in the range $1.5\leq 
        p_{max}\leq 2$. The black points correspond to three initial conditions of 
        orbits, namely $(x_0,y_0,z_0,t_0)=$ $(1.67209489949749,0,$
        $0.809297781072027,0)$ (point (1)), $(1.67192854271357,0,$ 
        $0.809236608877722,0)$ (point (2)),  $(1.67203944723618,0,$
        $0.809286279229481,0)$ (point (3)), $(1.67203675,0,$ $0.809286625,0)$ 
        (point (4)), $(1.6720387,0,$ $0.80928565,0)$ (point (5)).
        }
        \label{fig:gridfm}
\end{figure}
Numerical plots of the Arnold web on the action space can be obtained by 
plotting on a color or gray scale the value of a chaotic indicator as a 
function of the initial conditions of the orbits on the action space. 
Plots of this type, for different multidimensional Hamiltonian systems 
or mappings, were given, using different indicators 
\cite{Laskar01,Kaneko01,Froes02,Froes03,Cincot03,Giordano01}.
Here we use the APLE index $p_{max}$, for $T=10^4$ periods, in order to 
produce a plot of the Arnold web for the mapping (\ref{frm}), $\epsilon =0.05 $.
We set the initial values of the angles $ y_0 = t_0 =0$ and use a 
$500 \times 500$ grid  of initial conditions on the action plane $(x_0,z_0)$.
For all these orbits the initial deviation vector is $(dx_0,dy_0,dz_0,dt_0) = 
(\frac{1}{\sqrt{2}},0,0,\frac{1}{\sqrt{2}}) $. The results are shown in 
figure \ref{fig:gridfm}a. The color scale corresponds 
to values of $p_{max}$ in the range $0.6\leq  p_{max}\leq 1.4$, while, if 
the calculated value of $p_{max}$ for an orbit is outside the above interval, 
the associated color in the plot is replaced by that corresponding to the 
lower limit 0.6 or the upper limit 1.4.

\begin{figure}
	\centerline{\includegraphics[width=40pc] {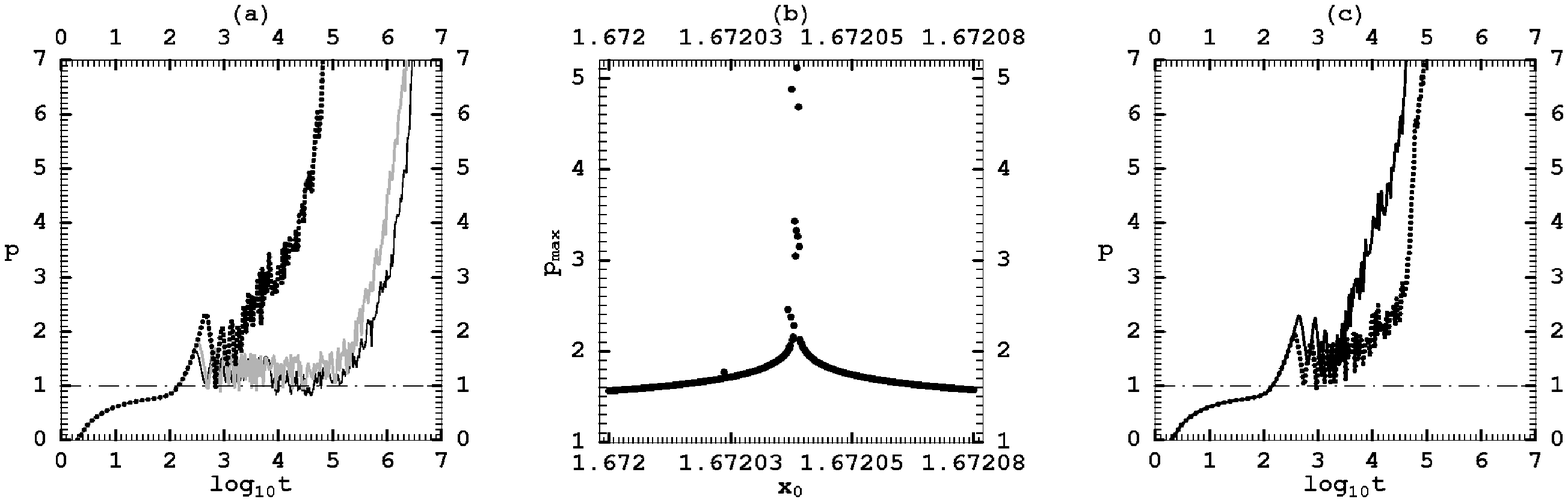}} 
         \caption{
         (a) The time evolution of the APLE for the three orbits 
         with initial conditions as in figure \ref{fig:gridfm}b (thick dotted black curve for 
         the point (1), gray curve for the point (2), and thin black for the point (3) 
         point). (b) The value of $p_{max}$ after $T=10^4$ iterations as a function of 
         the initial condition $x_0$ of a set of orbits in a segment crossing 
         perpendicularly the resonance line of figure \ref{fig:gridfm}a. 
         (c) Time evolution of $p$ for two orbits in the same line, with initial conditions 
         as in figure \ref{fig:gridfm}b. The thick dotted line corresponds to the point (4) 
         (orbit near the border of the resonance chaotic layer) and the thin continuous 
         line to the point (5) (orbit near the center of the resonance chaotic 
         layer).
         }
	\label{fig:resorb}
\end{figure}

The web of resonances is clearly visible in the plot of figure \ref{fig:gridfm}a, 
which is similar to plots of the same system produced in \cite{Froes03} 
using a different indicator, namely the FLI. Now, these authors 
studied in detail the diffusion in the action space of chaotic orbits 
starting at the chaotic border of a single resonance domain. When $\epsilon$ 
is sufficiently small, the system is in the `Nekhoroshev regime' in which the 
diffusion coefficient is found to be exponentially small in the inverse of 
the perturbation, i.e., $D\propto \exp\big(-(\epsilon_0/\epsilon)^b\big)$ 
for some positive exponent $b$ (estimated as $b=0.28$ by the same authors). 
Since the diffusing orbits are in general very weakly chaotic, we expect 
that some of them exhibit the metastable behavior associated with a constant 
rate of production of the Tsallis q-entropy. We find this to be the 
case for some of the orbits explored in \cite{Froes03}. If a 
zoom of figure \ref{fig:gridfm}a is made in the neighborhood of a chaotic 
border of the resonance $x-2z=0$ (figure \ref{fig:gridfm}b), the orbits 
studied in \cite{Froes03} (their figure 5) correspond to the initial conditions
\newline
$(x_0,y_0)=(1.67209489949749,0,0.809297781072027,0)$ (point (1)) or
\newline
$(1.67192854271357,0,0.809236608877722,0)$ (point (2)) and\\
$ (1.67203944723618,0,0.809286279229481,0) $ (point (3)). The point 
(3) is near a line passing from the center of the chaotic zone of figure 
\ref{fig:gridfm}b, while the two other points are near the edge of the 
same zone. Figure \ref{fig:resorb}a then shows the time evolution of the 
APLE, $p$ vs. $ \log_{10} t$ for these orbits. Clearly, in the case of 
the point (1) in figure \ref{fig:gridfm}b, the resulting orbits shows no 
visible metastable behavior, while such a behavior is clearly exhibited 
by the two orbits with initial conditions on the edge of the chaotic 
border. In fact, the calculation of the Lyapunov exponents $\lambda$ for 
these orbits \cite{Froes03} shows that the exponent of the 
first orbit is of order $\lambda\sim 10^{-2}$, while it is two orders of 
magnitude smaller $\lambda\sim 10^{-4}$ for the orbits near the edge 
of the chaotic border. We conclude that even if a system is in the 
`Nekhoroshev regime', not all the chaotic orbits manifest the metastable 
behavior associated with the constant rate of production of the q-entropy, 
but only the orbits with initial conditions close to the edge of chaotic 
borders separating resonance from non-resonance domains. 
\begin{figure}
	\centerline{\includegraphics[width=20pc] {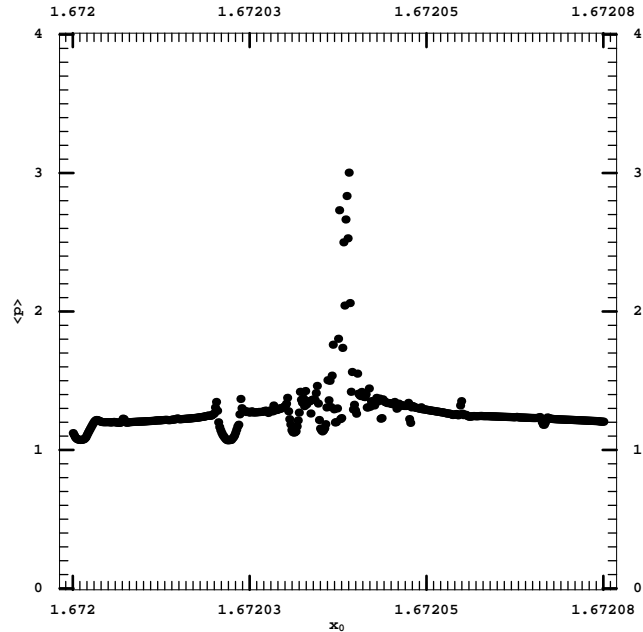}} 
        \caption{
         As in \ref{fig:gridfm}a, but for the average value $<p>$ 
         calculated over $10^4$ iterations per orbit. 
        }
	\label{fig:averpfm}
\end{figure}

In order to further substantiate this conclusion, figure \ref{fig:resorb}b 
shows the value of $p_{max}$ for $T=10^4$ versus the value of the coordinate 
$x_0$ by a detailed scanning along a line passing through the chaotic 
border of figure \ref{fig:gridfm}b in the direction perpendicular to the 
direction of the resonance under study. We see that as the center of the 
chaotic border is approached the value of $ p_{max} $ grows abruptly.
This plot is similar to the plot of $p_{max}$ for the crossing of a 
thin separatrix chaotic layer in the 2D standard map (figure \ref{fig:gridsm}a), 
a fact expected 
since the dynamics near the chaotic border of a single resonance domain 
is qualitatively given by a separatrix-like map \cite{Chirik01}. At any 
rate, figure \ref{fig:resorb}b clearly shows that a relatively small value 
of the APLE (above unity), leading to a long lasting metastable behavior 
of the orbits, can be expected only near the edge of the chaotic border 
of the resonance. In fact, if we take two initial conditions along the 
scanning line of figure \ref{fig:resorb}b (points (4) and (5) in figure 
\ref{fig:gridfm}b), the time interval of metastable behavior (figure 
\ref{fig:resorb}c) is longer for the orbit (4), which is closer to the 
border of the chaotic layer than for the orbit (5) which is closer to 
the center of the layer. However, both orbits are relatively closer to 
the center than the orbits (1) and (2), and, consequently, they both have 
shorter metastable time intervals than the latter orbits (figure 
\ref{fig:resorb}a).  

Finally, figure \ref{fig:averpfm} shows the average value 
$<p>$ over an interval $0\leq t\leq 10^4$ for all the orbits with initial 
conditions in the same interval as in figure \ref{fig:resorb}b. These orbits 
have a value of $<p>$ always above unity, and the mean of all values in the 
interval is $\overline{<p>}=1.27$, yielding an entropic index $\bar{q}=0.21$.

\section{Conclusions}
We studied the connection between the production of Tsallis q-entropy and the 
behavior of the variational equations of motion for weakly chaotic orbits in 
conservative dynamical systems. Our main findings can be summarized as follows:

1) The solutions of the variational equations present long transient time 
intervals in which the length of the deviation vector increases almost as 
a power-law. This allows to define an almost constant average power-law 
exponent (APLE) during the whole transient interval. 

2) This `metastable' behavior can be justified theoretically by showing that 
it is caused by the growth of the deviation vectors inside separatrix-like 
thin chaotic layers, which is of the form $\xi(t)\approx at +e^{\lambda t}$, 
with $a>>1$ and $\lambda<<1$. The latter law appears almost as a power law 
$\xi(t)\approx t^p$ for time intervals up to $t_{2c}\approx \lambda^{-1}$. 

3) The average value of the APLE in a thin chaotic layer can be used to 
determine an average value of the q-entropic index for which the Tsallis 
entropy exhibits a constant rate of increase.

4) The APLE can be used as an efficient numerical indicator distinguishing 
regular from weakly chaotic orbits. In that respect, it is equally powerful 
to other established indicators, as the FLI or the MEGNO. The advantage of 
the APLE is that it gives also the average value of the q-entropic index 
within a weak chaotic layer. 

5) Numerical implementations of the APLE are given in low-dimensional 
symplectic mappings. The APLE is calculated in a thin chaotic layer and 
in the stickiness region of an island of stability in the 2D standard map. 
We then use it in order to visualize the Arnold web of resonances in a 
4D map, and calculate the q-entropic index at the chaotic border of a 
single resonance domain of the same map. 

{\bf Acknowledgements:} G. Lukes-Gerakopoulos was supported in part by 
the Greek Foundation of State Scholarships (IKY) and by the Research 
Committee of the Academy of Athens. We thank Prof. G. Contopoulos for 
useful suggestions and a careful reading of the manuscript.

\end{document}